\documentclass{elsart}

\usepackage[dvips]{color}
\usepackage{textcomp}
\usepackage{amsmath}
\usepackage{amssymb}
\usepackage{graphicx}
\usepackage{wrapfig}
\usepackage[ps2pdf,a4paper=false,letterpaper=true,colorlinks=true,urlcolor=blue,citecolor=magenta,bookmarks=false]{hyperref}
\input epsf

\begin{document}

\begin{frontmatter}

\title{Search for \emph{T} Violation in Charm Meson Decays}

\collaboration{The~FOCUS~Collaboration}\footnotemark
\author[ucd]{J.~M.~Link}
\author[ucd]{P.~M.~Yager}
\author[cbpf]{J.~C.~Anjos}
\author[cbpf]{I.~Bediaga}
\author[cbpf]{C.~Castromonte}
\author[cbpf]{A.~A.~Machado}
\author[cbpf]{J.~Magnin}
\author[cbpf]{A.~Massafferri}
\author[cbpf]{J.~M.~de~Miranda}
\author[cbpf]{I.~M.~Pepe}
% ...

\address[ucd]{University of California, Davis, CA 95616}
\address[cbpf]{Centro Brasileiro de Pesquisas F\'\i sicas, Rio de Janeiro, RJ, Brazil}
% ...

\footnotetext{See \textrm{http://www-focus.fnal.gov/authors.html} for additional author information.}

\begin{abstract}
Using data from the FOCUS (E831) experiment,  we have searched for 
\emph{T} violation in charm meson decays using the four-body decay channels 
\hbox{$D^0 \to K^-K^+\pi^-\pi^+$}, \hbox{$D^+ \to K^0_SK^+\pi^-\pi^+$}, and 
\hbox{$D^+_s \to K^0_SK^+\pi^-\pi^+$}. The \emph{T} violation asymmetry is obtained using triple-product 
correlations and assuming the validity of the \emph{CPT} theorem. We find the asymmetry values to be  
 ${A_\mathrm{Tviol} (D^0) } = 0.010 \pm 0.057(\mathrm{stat.}) \pm 0.037(\mathrm{syst.})$,
  ${A_\mathrm{Tviol} (D^+) } = 0.023 \pm 0.062(\mathrm{stat.}) \pm 0.022(\mathrm{syst.})$, and
  ${A_\mathrm{Tviol} (D^+_s) } = -0.036 \pm 0.067(\mathrm{stat.}) \pm 0.023(\mathrm{syst.})$.  Each
measurement is consistent with no \emph{T} violation. New measurements of the
\emph{CP} asymmetries for some of these decay modes are also presented.

\end{abstract}

\end{frontmatter}  

\textbf{1. Introduction}

 The origin of \emph{CP} violation remains one of the most important open questions in particle physics. 
Within the Standard Model, \emph{CP} violation arises due to the presence of a phase in the
Cabibbo-Kobayashi-Maskawa (CKM) quark mixing matrix. Although the main focus has been on
rate asymmetries, there is another type of \emph{CP} violating signal which could potentially
reveal the presence of physics beyond the Standard Model. Triple-product correlations of the
form $\vec{v_1}\cdot(\vec{v_2}\times\vec{v_3})$, where each $\vec{v_i}$ is a spin or momentum,
are odd under time reversal (\emph{T}). By the \emph{CPT} theorem, a nonzero value for these 
correlations would also be a signal 
of \emph{CP} violation. A nonzero triple-product correlation is evidenced by a nonzero value 
of the asymmetry~\cite{Bensalem}

\begin{equation}
 A_{T}  \equiv \frac{\Gamma(\vec{v_1}\cdot(\vec{v_2}\times\vec{v_3})>0) - 
                     \Gamma(\vec{v_1}\cdot(\vec{v_2}\times\vec{v_3})<0)}
		    {\Gamma(\vec{v_1}\cdot(\vec{v_2}\times\vec{v_3})>0) + 
                     \Gamma(\vec{v_1}\cdot(\vec{v_2}\times\vec{v_3})<0)}   
\end{equation}

where $\Gamma$ is the decay rate for the process. There is a well-known technical
complication: strong phases can produce a nonzero value of $A_{T}$, even if the weak phases
are zero, that is \emph{CP} and \emph{T}  violation are not necessarily present. 
Thus, strictly speaking, the
asymmetry $A_{T}$ is not in fact a \emph{T}-violating effect. Nevertheless, one can still obtain
a true \emph{T}-violating signal by measuring a nonzero value of

\begin{equation}
 A_\mathrm{Tviol}  \equiv \frac{1}{2}(A_{T}-\overline{A_{T}}) 
\end{equation}

where $\overline{A_{T}}$ is the \emph{T}-odd asymmetry measured in the \emph{CP}-conjugate decay 
process~\cite{Bensalem2}.

 This study was inspired by a paper of Ikaros Bigi~\cite{Bigi}. In this paper Bigi 
suggested a search for \emph{T} violation by looking at the triple-product correlation 
(using the momenta of the final state particles) in the decay mode 
\hbox{$D^0 \to K^-K^+\pi^-\pi^+$}. Such a correlation must necessarily involve at 
least four final-state particles. This can be understood by considering the rest frame 
of the decaying particle and invoking momentum conservation. The number of independent 
three-momenta is one less than the number of final-state particles, so a triple product 
composed entirely of momenta requires four particles in the final state~\cite{Valencia}.

We calculate $A_\mathrm{Tviol}$ for the decay 
modes $D^0 \to K^-K^+\pi^-\pi^+$ and $D^+_{(s)} \to K^0_S K^+\pi^-\pi^+$ using 
data from the FOCUS experiment. 

 FOCUS is a charm photoproduction experiment~\cite{spectro} which collected 
data during the 1996--97 fixed target run at Fermilab. Electron and positron beams (with
typically $300~\textrm{GeV}$ endpoint energy) obtained from the $800~\textrm{GeV}$ Tevatron
proton beam produce, by means of bremsstrahlung, a photon beam which
interacts with a segmented BeO target. The mean photon energy for triggered
events is $\sim 180~\textrm{GeV}$. A system of three multicell threshold \v{C}erenkov
counters performs the charged particle identification, separating kaons from
pions up to $60~\textrm{GeV}/c$ of momentum. Two systems of silicon microvertex
detectors are used to track particles: the first system consists of 4 planes
of microstrips interleaved with the experimental target~\cite{WJohns} and the
second system consists of 12 planes of microstrips located downstream of the
target. These detectors provide high resolution in the transverse plane
(approximately $9~\mu\textrm{m}$), allowing the identification and separation of the 
primary (production) and the charm secondary (decay) vertices. Charged particle
momentum is determined by measuring deflections in two magnets of
opposite polarity through five stations of multiwire proportional chambers.

\vskip 0.5cm \textbf{2. Search for \emph{T} violation in the decay mode 
{\boldmath$D^0 \to K^-K^+\pi^-\pi^+$}}

 The decay mode $D^0 \to K^-K^+\pi^-\pi^+$ is Cabibbo-suppressed and may be produced as 
a non-resonant final state or via two-body and three-body intermediate resonant states.
In a previous paper we determined its resonant substructure and the branching ratio
$\Gamma(D^0 \to K^-K^+\pi^-\pi^+)/\Gamma(D^0 \to K^-\pi^-\pi^+\pi^+)$~\cite{Alberto}.

The final states are selected using a \textit{candidate driven vertex
algorithm}~\cite{spectro}. A secondary vertex is formed from the four
candidate tracks. The momentum vector of the resultant $D^{0}$ candidate is used as
a \textit{seed} track to intersect the other reconstructed tracks and to 
search for a primary vertex. The confidence levels of both vertices are
required to be greater than $1\%$. Once the production and decay vertices are 
determined, the distance $L$ between the vertices and its error $\sigma _{L}$ are computed. 
The quantity $L$\thinspace /\thinspace $\sigma _{L}$ is an unbiased measure of the significance of 
detachment between the primary and secondary vertices. This is the most important variable
for separating charm events from non-charm prompt backgrounds. Signal quality is further enhanced
by cutting on \emph{Iso2}, which is the confidence level that other tracks in the event might be associated
with the secondary vertex. We use $L$\thinspace /\thinspace $\sigma _{L}$ $>$ $6$ and 
\emph{Iso2} $<$ 10$\%$. We also require the $D^0$ momentum to be in the range
25--250 GeV/$c$ 
(a very loose cut) and the primary vertex to be formed with at least two reconstructed tracks 
in addition to the $D^0$ seed. 

The \v{C}erenkov identification cuts used in FOCUS are based on likelihood ratios 
between the various particle identification hypotheses. These likelihoods are computed 
for a given track from the observed firing response (on or off) of all the cells that are
within the track's ($\beta =1$) \v{C}erenkov cone for each of our three 
\v{C}erenkov counters. The product of all firing probabilities for all the cells
within the three \v{C}erenkov cones produces a $\chi ^{2}$-like variable 
$W_{i}=-2\ln (\mathrm{Likelihood})$ where $i$ ranges over the electron, pion,
kaon and proton hypotheses~\cite{cerenkov}. All kaon tracks are required
to have $\Delta _{K}=W_{\pi }-W_{K}$ (kaonicity) greater than $3$; whereas all
the pion tracks are required to be separated by less than $5$ units from the best
hypothesis, that is $picon=W_\mathrm{min}-W_{\pi }$ (pion consistency) is greater 
than $-$5.

In addition to these cuts (also used in our previous analysis of this decay mode), we 
require a $D^*$-tag. The sign of the bachelor pion in the  $D^{*\pm}$ decay chain 
$D^{*+(-)} \to D^0(\overline{D^0})\pi^{+(-)}$ is used to identify the neutral $D$ as either
a $D^0$ or a $\overline{D^0}$. We require that the mass difference between the $D^0$ and the
$D^{*}$ mass be within $4$ MeV/$c^2$ of the nominal mass difference~\cite{PDG}.

Using the set of selection cuts just described, we obtain the invariant mass distributions
for $K^{-}K^{+}\pi^{-}\pi^{+}$ shown in Fig.~1, where the first plot is the total sample
and the other two plots show the $D^0$ and $\overline{D^0}$ samples separately. 

The mass plots are fit with a function that includes two Gaussians with the same mean but 
different sigmas to take into account different momentum resolutions in our 
spectrometer~\cite{spectro} and a second-order polynomial for the combinatorial background.
A log-likelihood fit gives a signal of $828 \pm 46$ $K^{-}K^{+}\pi^{-}\pi^{+}$ events for 
the total sample, $362 \pm 31$ $D^0$ events, and $472 \pm 34$ $\overline{D^0}$ events.
The fitted $D^0$ masses are in good agreement with the world average~\cite{PDG} 
and the widths are in good agreement with those of our Monte Carlo simulation.

From the $D^0$ sample we can form a \emph{T}-odd correlation with the momenta:
\begin{equation}
C_T \equiv \vec{p}_{K^+}\cdot(\vec{p}_{\pi^+}\times \vec{p}_{\pi^-})
\end{equation}

and from the $\overline{D^0}$  sample we form:
\begin{equation} 
\overline{C_T} \equiv \vec{p}_{K^-}\cdot(\vec{p}_{\pi^-}\times \vec{p}_{\pi^+}).
\end{equation}

 As we have seen in the introduction, finding a distribution of $C_T$ different from 
$-\overline{C_T}$ establishes \emph{CP} violation~\cite{Bigi}.

Fig.~2 shows $D^0$($\overline{D^0}$) signals separated by the sign
of $C_T(\overline{C_T})$. A log-likelihood fit, with
the same fit function described previously, gives the yields summarized in  
Table~\ref{ctyields}.

\begin{table}[h!]
\begin{center}
\caption{$D^0$ ($\overline{D^0}$) yields split by $C_T(\overline{C_T})$ sign.}
\label{ctyields}
\begin{tabular}{|l|l|l|}
\hline
Decay mode & Request & Events \\ 
\hline
$D^0 \to K^-K^+\pi^-\pi^+$            & $C_T>0$            & $  174 \pm 21 $ \\
$D^0 \to K^-K^+\pi^-\pi^+$            & $C_T<0$            & $  190 \pm 24 $ \\ 
$\overline{D^0} \to K^-K^+\pi^-\pi^+$ & $\overline{C_T}>0$ & $  255 \pm 24 $ \\
$\overline{D^0} \to K^-K^+\pi^-\pi^+$ & $\overline{C_T}<0$ & $  220 \pm 25 $ \\ 
\hline
\end{tabular}
\end{center}
\end{table}

Before forming the asymmetry $A_{T}(\overline{A_{T}})$ 
we have to correct for detection efficiencies, accounting for possible
differences in spectrometer acceptance and \v{C}erenkov identification efficiency
for positive/negative kaons and pions.\footnote{It is well-known that in fixed-target experiments
there are production asymmetries between charm and anticharm
particles. As a result the $D^0$ momentum distribution is
different from the $\overline{D^0}$ distribution.}
This is, however, a small effect. From the efficiency corrected yields we compute the 
asymmetry:

\begin{equation}
 A_{T}  = \frac{\Gamma(C_T>0)-\Gamma(C_T<0)}{\Gamma(C_T>0)+\Gamma(C_T<0)}	       
\end{equation}

and

\begin{equation}
\overline{A_{T}} = \frac{\Gamma(-\overline{C_T}>0)-\Gamma(-\overline{C_T}<0)}
                        {\Gamma(-\overline{C_T}>0)+\Gamma(-\overline{C_T}<0)}.	       
\end{equation}

The resulting \emph{T}-violation asymmetry $A_\mathrm{Tviol}$ is:

\begin{equation}
 A_\mathrm{Tviol} = \frac{1}{2}(A_{T}-\overline{A_{T}}) = 0.010 \pm 0.057. 
\end{equation}

Without the efficiency correction it would have been ${A_\mathrm{Tviol}} = 0.014 \pm 0.057$.
 
 This determination has been tested by modifying each of the vertex
and \v{C}erenkov cuts individually. Although the statistics is limited, 
the \emph{T}-violation asymmetry is 
stable versus several sets of cuts as shown in Fig.~\ref{ATvscuts}.
All the measurements are consistent with $0$
for the \emph{T}-violation asymmetry.

\vskip 0.5cm \textbf{3. Search for \emph{T} violation in the decay mode 
{\boldmath$D \to K^0_S K^+\pi^-\pi^+$}}
 
The decay channel $D^+ \to K^0_S K^+\pi^-\pi^+$ is Cabibbo-suppressed and like $D^0 \to K^-
K^+\pi^-\pi^+$,
it may be produced as
a non-resonant final state or via two-body and three-body intermediate resonant states.
Its relative branching ratio
$\Gamma(D^+ \to K^0_SK^+\pi^-\pi^+)/\Gamma(D^+ \to K^0_S\pi^-\pi^+\pi^+)$ has been 
measured~\cite{Ko}.
$D^+_s \to K^0_SK^+\pi^-\pi^+$ is observed in the same histogram as $D^+ \to
K^0_S K^+\pi^-\pi^+$
and we fit for both signals.

The final states are selected using a \textit{candidate driven vertex
algorithm} as described in the previous section. 
The $K^0_S$ is reconstructed using techniques described
elsewhere~\cite{Cumalat}. The $K^0_S$ and the charged
tracks are used to form a $D$ candidate which is used as
a \textit{seed} track to intersect the other reconstructed tracks and to
search for a primary vertex. The confidence levels of both vertices must
be greater than $1\%$. 
We also use $L$\thinspace /\thinspace $\sigma _{L}$ $>$ $6$
and
\emph{Iso2} $< 1 \%$ and require 
the primary vertex to be composed of at least two reconstructed tracks
in addition to the $D$ seed.

Using these selection cuts, we obtain the invariant mass distributions
for $K^{0}_SK^{+}\pi^{-}\pi^{+}$ shown in Fig.~4, where the top plot is the total sample
and the bottom two plots show the $D$ and $\overline{D}$ samples separately.

The mass plots are fit with a function that includes a  Gaussian for the $D^+$ and 
a Gaussian for the $D^+_s$
with the widths fixed to those given from our Monte Carlo simulations. We 
use a second-order polynomial for the combinatorial background in addition to two
reflection peaks from $\Lambda^+_c \to p K^{0}_S\pi^{-}\pi^{+}$ and 
$D^+ \to K^0_S\pi^+\pi^{-}\pi^{+}$.  The $\Lambda^+_c$ yield is fixed after first fitting
the sample with the $K^+$ mass changed to the proton mass. The $D^+ \to K^0_S\pi^+\pi^{-}\pi^{+}$
yield is determined  by using the Monte Carlo misidentification rate of a pion as a kaon
and the yield of $D^+ \to K^0_SK^+\pi^{-}\pi^{+}$.
A log-likelihood fit gives a signal of $523 \pm 32$ events for
the $D^{\pm}$ and a signal of $508 \pm 34$ events for $D^{\pm}_s$.
The $K^{0}_SK^{+}\pi^{-}\pi^{+}$ sample has $240 \pm 22$ $D^+$  and $270 \pm 25$ $D^+_s$ events,
while the $K^{0}_SK^{-}\pi^{-}\pi^{+}$ sample has $282 \pm 23$ $D^-$  and $239 \pm 24$ $D^-_s$ events. 
The fitted $D$ masses are in good agreement with the world average~\cite{PDG}. Also the 
excess of $D^-$ over $D^+$ events is consistent with  more $\overline{D^0}$ mesons than $D^0$ 
mesons being produced.  These photoproduced excesses have been observed in previous higher 
statistics studies by FOCUS~\cite{CPV,Brian}.

The mass plots shown in  Fig.~5 are the $D^+_{(s)}$ ($D^-_{(s)}$) signals split by the sign
of $C_T(\overline{C_T})$. A log-likelihood fit, with
the same fit function described previously, gives the yields summarized in
Table~\ref{d+ctyields}.

\begin{table}[h!]
\begin{center}
\caption{$D^+_{(s)}$ ($D^-_{(s)}$) yields split by $C_T(\overline{C_T})$ sign.}
\label{d+ctyields}
\begin{tabular}{|l|l|l|l|}
\hline
Final State  & Request & $D^+$ Events & $D^+_s$ Events \\
\hline
$K^0_SK^+\pi^-\pi^+$            & $C_T>0$            & $  122 \pm 16 $ & $  126 \pm 17 $\\
$K^0_SK^+\pi^-\pi^+$            & $C_T<0$            & $  118 \pm 16 $ & $  147 \pm 18 $ \\
$K^0_SK^-\pi^-\pi^+$ & $\overline{C_T}>0$ & $  145 \pm 16 $ & $  120 \pm 17 $\\
$K^0_SK^-\pi^-\pi^+$ & $\overline{C_T}<0$ & $  137 \pm 16 $ & $  119 \pm 16 $\\
\hline
\end{tabular}
\end{center}
\end{table}

After correcting for detection and reconstruction efficiencies as given by the 
Monte Carlo simulation, we form the asymmetry  $A_{T}(\overline{A_{T}})$ as
given by equation (5) and equation (6):

\begin{equation}
 A_\mathrm{Tviol}(D^+) = \frac{1}{2}(A_{T}-\overline{A_{T}}) = 0.023 \pm 0.062
\end{equation}

\begin{equation}
 A_\mathrm{Tviol}(D^+_s) = \frac{1}{2}(A_{T}-\overline{A_{T}}) = -0.036 \pm 0.067.
\end{equation}

Without the efficiency corrections the numbers are essentially the same.  A scan of
$A_\mathrm{Tviol}$  under a variety of different selection criteria is presented in Fig.~6. 
%

%
%

\vskip 0.5cm \textbf{4. Systematic Uncertainties}

Systematic uncertainties on the \emph{T}-violation asymmetry  measurement can come from different sources.
We determine five independent contributions to the
systematic uncertainty: the \emph{split sample} component, the \emph{fit
variant} component, the component due to the particular choice of the
vertex and \v{C}erenkov cuts (discussed previously), the dilution due to an erroneous
$D^*$ tag for the $D^0 \to K^-K^+\pi^-\pi^+$ channel, and a component due to 
the limited statistics of the Monte Carlo.

The \emph{split sample} component addresses
the systematics
introduced by a residual difference between data and Monte Carlo, due to a
possible mismatch in the reproduction of the $D$
momentum and the changing experimental conditions during
data collection. This component has been determined by splitting data
into four independent subsamples, according to the $D$
momentum range (high and low momentum) and the configuration of the vertex detector,
that is, before and after the insertion of an upstream silicon system. A technique,
employed in FOCUS and in the predecessor experiment E687, modeled after the
\emph{S-factor method} from the Particle Data Group~\cite{PDG}, is used
to try to separate true systematic variations from statistical
fluctuations. The \emph{T}-violation asymmetry is evaluated for each of the $4~(=2^{2})$
statistically independent subsamples and a \emph{scaled variance} $\tilde{\sigma}$ (that
is the errors are boosted when $\chi ^{2}/(N-1)>1$) is calculated.
The \emph{split sample} variance $\sigma_\textrm{split}$ is defined as the
difference between the reported statistical variance and the scaled variance,
if the scaled variance exceeds the statistical variance~\cite{brkkpipi}.

Another possible source of systematic uncertainty is the \emph{fit variant}.
This component is computed by varying, in a reasonable manner, the fitting
conditions on the whole data set. In our study of the $D^0$ mode, we fixed the widths of the
Gaussians to the values obtained by the Monte Carlo simulation, we changed
the background parametrization (varying the degree of the polynomial), we modified
the fit function in order to take into account the reflection peak from
$D^0 \to K^-\pi^+\pi^-\pi^+$~\cite{Alberto}, and we use one Gaussian instead of two.
For all modes,  the variation of the computed efficiencies due to the different
resonant substructure simulated in the Monte Carlo has been taken into account.
The \emph{T}-violation values obtained by these variants are all {\em a priori} equally likely,
therefore this uncertainty can be estimated by the {\it r.m.s.}\ of the measurements~\cite{brkkpipi}.

Analogously to the \emph{fit variant}, the cut component is estimated using
the standard deviation of the several sets of cuts shown in Fig.~\ref{ATvscuts}
and Fig.~6.
Actually, this is an overestimate of the cut component because the statistics of
the cut samples are different.

An erroneous $D^*$ tag can obviously dilute the measured asymmetry $A_\mathrm{Tviol}$.
We find a dilution\footnote{The dilution, measured by means of our Monte
Carlo simulation, is defined as $D = \frac{R-W}{R+W}$. For the $D^0$ ($\overline{D^0}$) 
dilution, $R$ is the number of generated $D^0$ ($\overline{D^0}$) events  reconstructed 
correctly as $D^0$ ($\overline{D^0}$) and $W$ 
is the number of generated $D^0$ ($\overline{D^0}$) events reconstructed 
as $\overline{D^0}$ ($D^0$).} 
of $0.9846\pm0.0029$ for the $D^0$ sample and $0.9882\pm0.0025$ for the
$\overline{D^0}$ events (the dilutions are slightly different because, as we have already
seen, the $D^0$ and $\overline{D^0}$ momentum distributions are different). Then we
computed the $A_\mathrm{Tviol}$ asymmetry taking into account this dilution and estimated
the uncertainty by using the difference between this determination and the standard one.

 Finally, there is a further contribution due to the limited statistics of
the Monte Carlo simulation used to determine the efficiencies. Adding in
quadrature all of these components, we obtain the final systematic
errors which are summarized in Table~\ref{err_sist}.

\begin{table}[h!]
\begin{center}
\caption{Contribution to the systematic uncertainties of the \emph{T}-violation parameters for $D^0$, $D^+$, 
and $D^+_s$.}
\label{err_sist}
\begin{tabular}{|l|c|c|c|}
\hline
{Source}        & {$D^0$ Uncertainty} & {$D^+$ Uncertainty} & {$D^+_s$ Uncertainty}     \\
\hline
{Split sample}  &  0.000  &  0.000 &  0.000 \\
{Fit Variant}   &  0.009  &  0.006 &  0.004 \\
{Set of cuts}   &  0.035  &  0.021 &  0.022 \\
{$D^*$-tag dilution} & 0.002 & --- &  ---     \\
{MC statistics} &  0.009  &  0.004 &  0.006 \\ \hline
{Total systematic error} &  0.037  &  0.022 & 0.023\\
\hline
\end{tabular}
\end{center}
\end{table}

%
%

\vskip 0.5cm \textbf{5. Conclusions}

 Using data from the FOCUS (E831) experiment at Fermilab we have searched for 
\emph{T} violation in charm meson decays.   It is a clean and
     alternative way to search for \emph{CP} violation. This is the first time such a
measurement has been performed in the charm sector.

We determine the final values for the \emph{T}-violation asymmetries to be 

  ${A_\mathrm{Tviol} (D^0) } = 0.010 \pm 0.057(\mathrm{stat.}) \pm 0.037(\mathrm{syst.})$,

  ${A_\mathrm{Tviol} (D^+) } = 0.023 \pm 0.062(\mathrm{stat.}) \pm 0.022(\mathrm{syst.})$, and

  ${A_\mathrm{Tviol} (D^+_s) } = -0.036 \pm 0.067(\mathrm{stat.}) \pm 0.023(\mathrm{syst.})$.

It is interesting to compare the $A_\mathrm{Tviol}$ measurements with the usual
 \emph{CP} asymmetry measurements. Following the procedure described in a previous
 paper~\cite{CPV} we determine $A_{CP}$ for $D^0 \to K^-K^+\pi^-\pi^+$ and
 $D^+ \to K^0_SK^+\pi^-\pi^+$, where we used $D^0 \to K^-\pi^+\pi^-\pi^+$ and
 $D^+ \to K^0_S\pi^+\pi^-\pi^+$ to account for differences at production level.
Systematic errors are obtained from the same 
sources and in the same manner as for the $A_{Tviol}$ measurement.

  We measure:

   ${A_{CP} (D^0) } = -0.082 \pm 0.056(\mathrm{stat.}) \pm 0.047(\mathrm{syst.})$ and 

   ${A_{CP} (D^+) } = -0.042 \pm 0.064(\mathrm{stat.}) \pm 0.022(\mathrm{syst.})$.

Both $A_\mathrm{Tviol}$ and $A_{CP}$ are consistent with zero.
While our measurements are consistent with no \emph{T} 
violation, we encourage higher statistics experiments to repeat these measurements. 

%
%

\vspace{0.5cm} \textbf{6. Acknowledgements}

We want to thank Prof. Ikaros Bigi for illuminating discussions
on the \emph{T}-odd correlation subject.

Further, we wish to acknowledge the assistance of the staffs of Fermi National
Accelerator Laboratory, the INFN of Italy, and the physics departments of
the collaborating institutions. This research was supported in part by the
U.~S. National Science Foundation, the U.~S. Department of Energy, the
Italian Istituto Nazionale di Fisica Nucleare and Ministero della Istruzione
Universit\`a e Ricerca, the Brazilian Conselho Nacional de Desenvolvimento
Cient\'{\i}fico e Tecnol\'ogico, CONACyT-M\'exico, and the Korea Research
Foundation of the Korean Ministry of Education.

%
%

%
%

\newpage
\begin{figure}[!!t]
\begin{center}
\includegraphics[width=13.0cm]{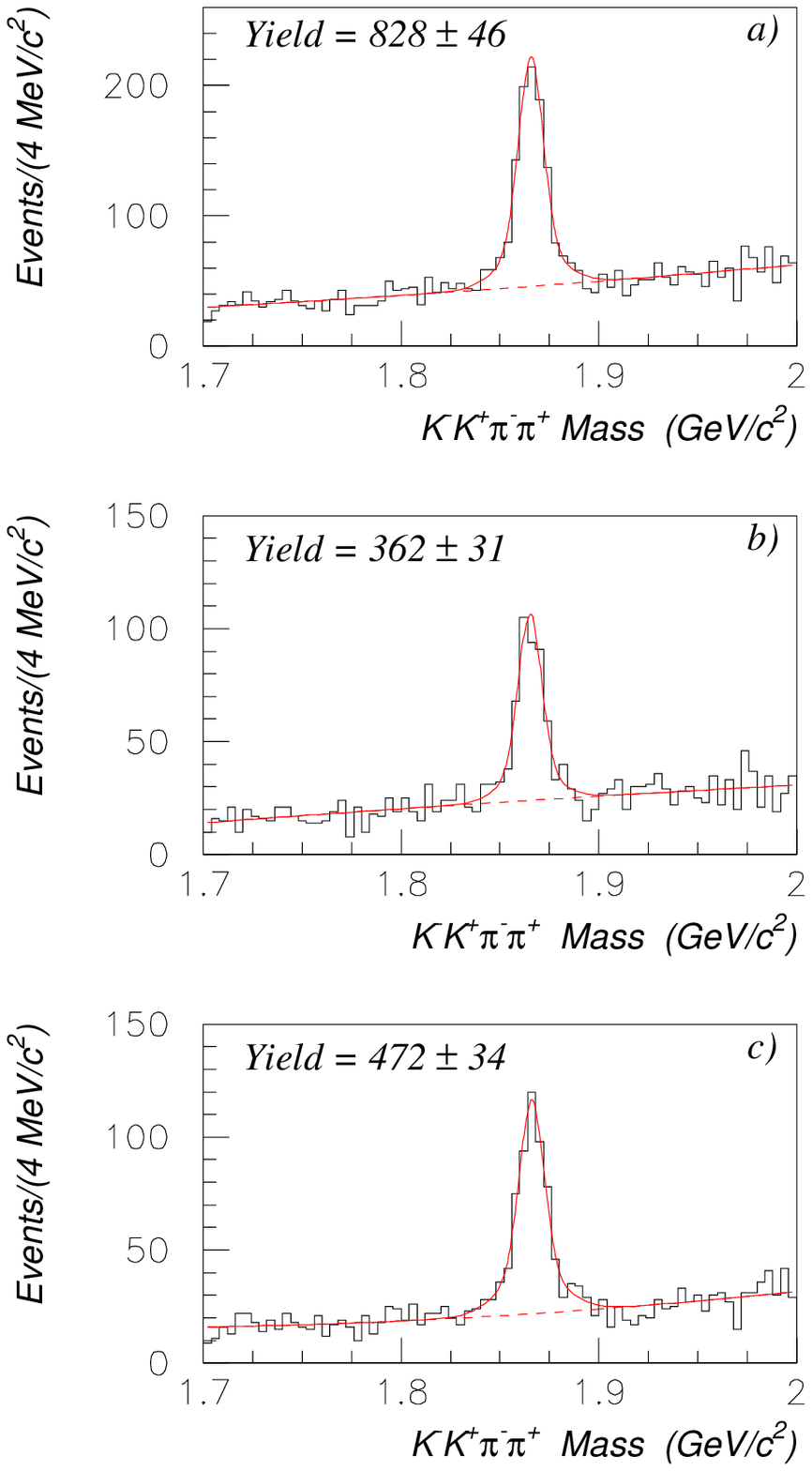}
\end{center}
\caption{$K^-K^+\protect\pi^-\protect\pi^+$ invariant mass distributions 
for: (a) total sample $D^{*+(-)} \to D^0(\overline{D^0})\pi^{+(-)}$,
(b) $D^0$ sample, $D^{*+} \to D^0\pi^{+}$ and (c) $\overline{D^0}$ sample, 
$D^{*-} \to \overline{D^0}\pi^{-}$. The fit (solid curve) is explained in the text.}
\label{d*masses}
\end{figure}

\newpage
\begin{figure}[!!t]
\begin{center}
\includegraphics[width=15.0cm]{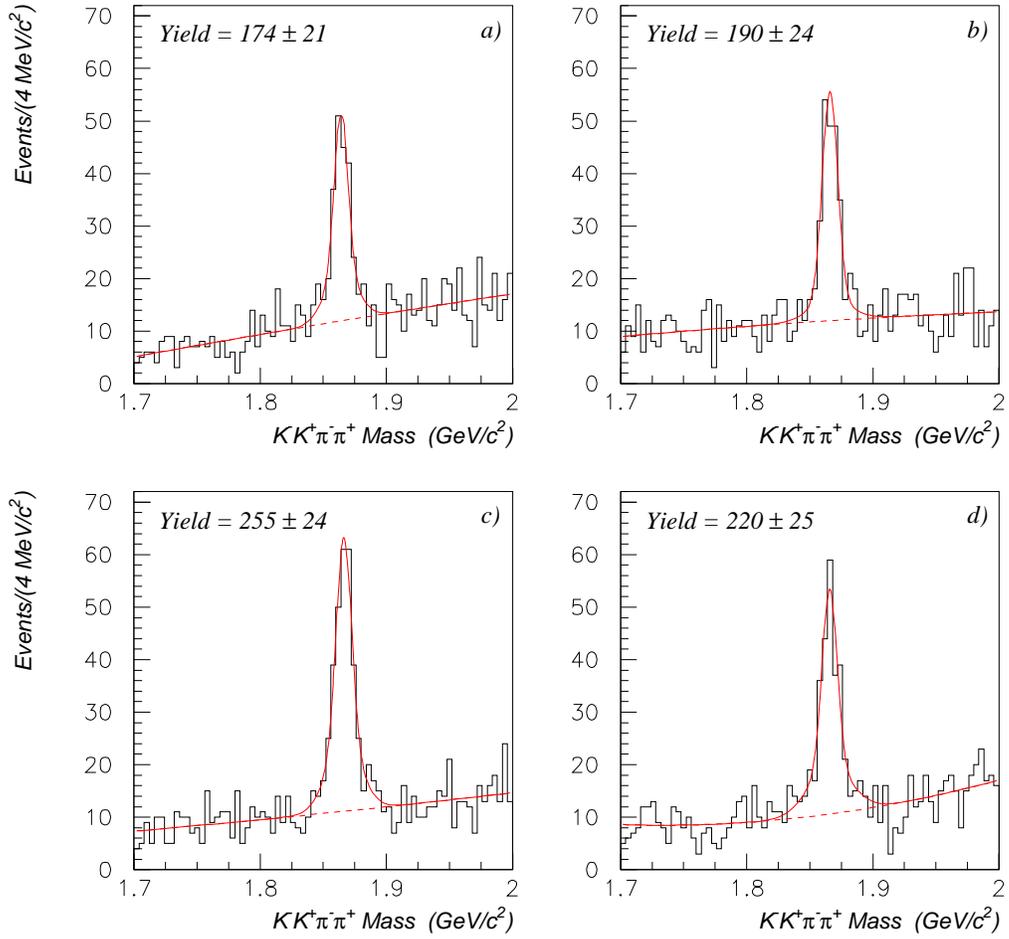}
\end{center}
\caption{$K^-K^+\protect\pi^-\protect\pi^+$ invariant mass distributions for: 
(a) $D^0$ sample with $C_T>0$, (b) $D^0$ sample with $C_T<0$,
(c) $\overline{D^0}$ sample with $\overline{C_T}>0$ and 
(d) $\overline{D^0}$ sample with $\overline{C_T}<0$.
The fit (solid curve) is explained in the text.}
\label{toddmasses}
\end{figure}

\newpage
\begin{figure}[!!t]
\begin{center}
\includegraphics[width=15.0cm]{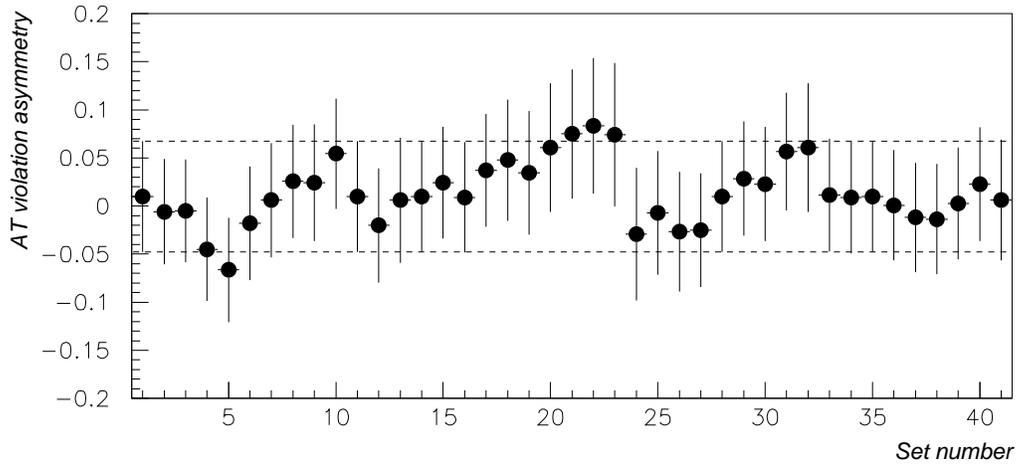}
\end{center}
\caption{\emph{T}-violating asymmetry $A_\mathrm{Tviol}$
versus several sets of cuts. We varied the confidence level of the secondary vertex from 
$1\%$ to $5\%$ (5 points), \emph{Iso2} from $10^{-6}$ to $1$ (7 points), 
$L$\thinspace /\thinspace $\sigma _{L}$ from $5$ to $15$ (11 points), 
$\Delta _{K}$ from $1$ to $5$ (9 points), {\it picon} from $-6$ to $-2$ (9 points). 
The dashed lines show the quoted $A_\mathrm{Tviol}$ asymmetry $\pm 1 \sigma$.}
\label{ATvscuts}
\end{figure}

\begin{figure}[htb!]
\begin{center}
\includegraphics[width=7.5cm]{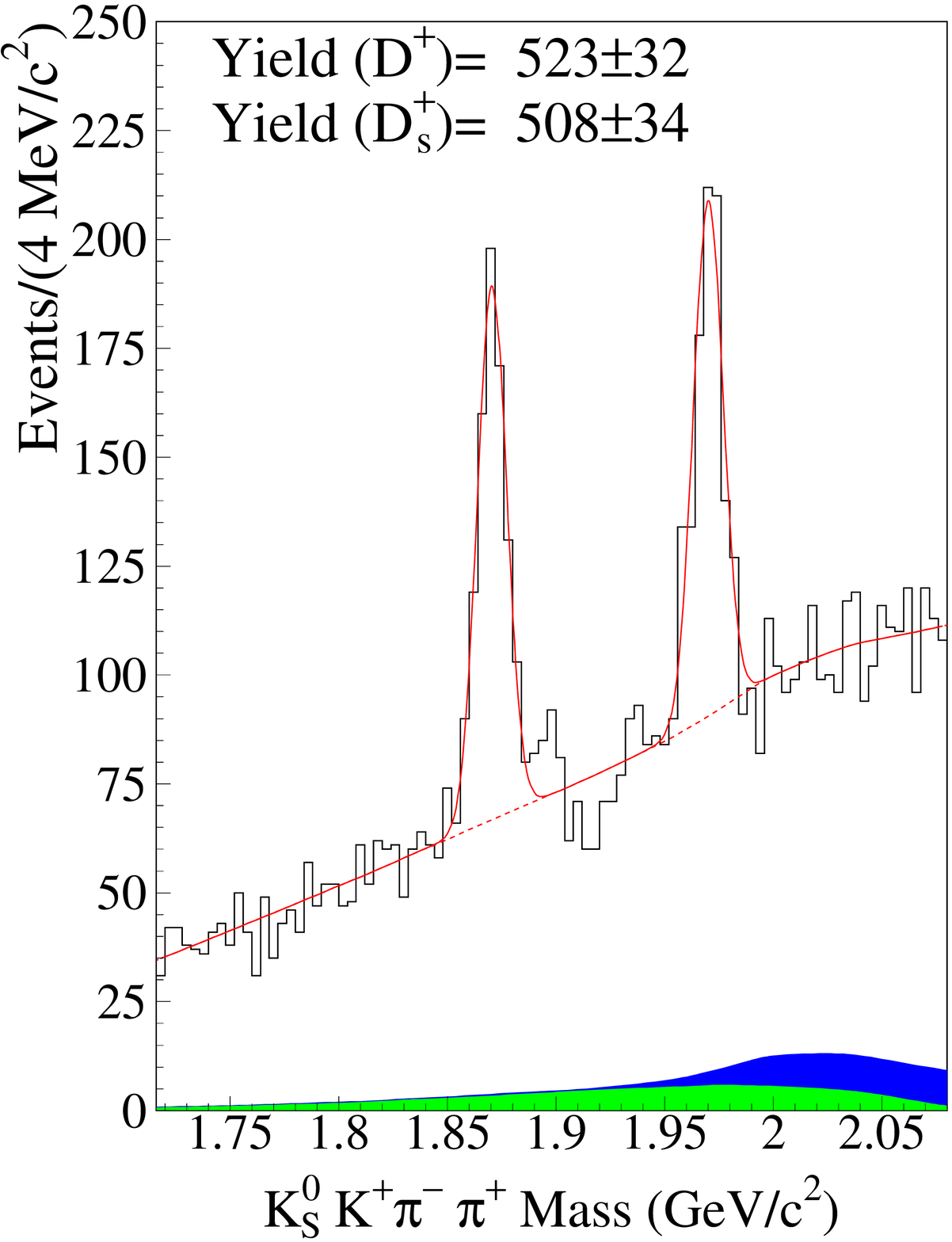}
\includegraphics[width=6.5cm]{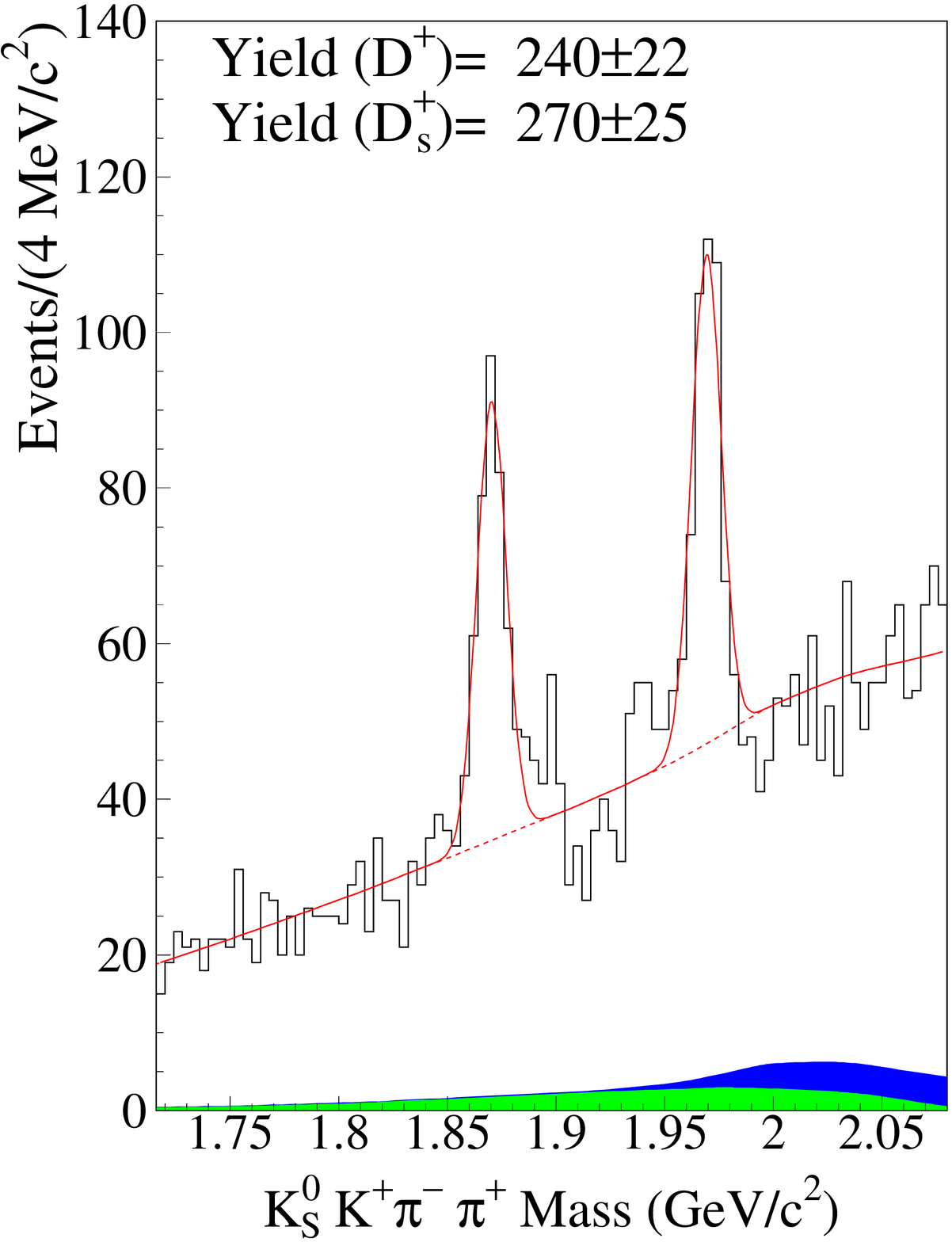}
\includegraphics[width=6.5cm]{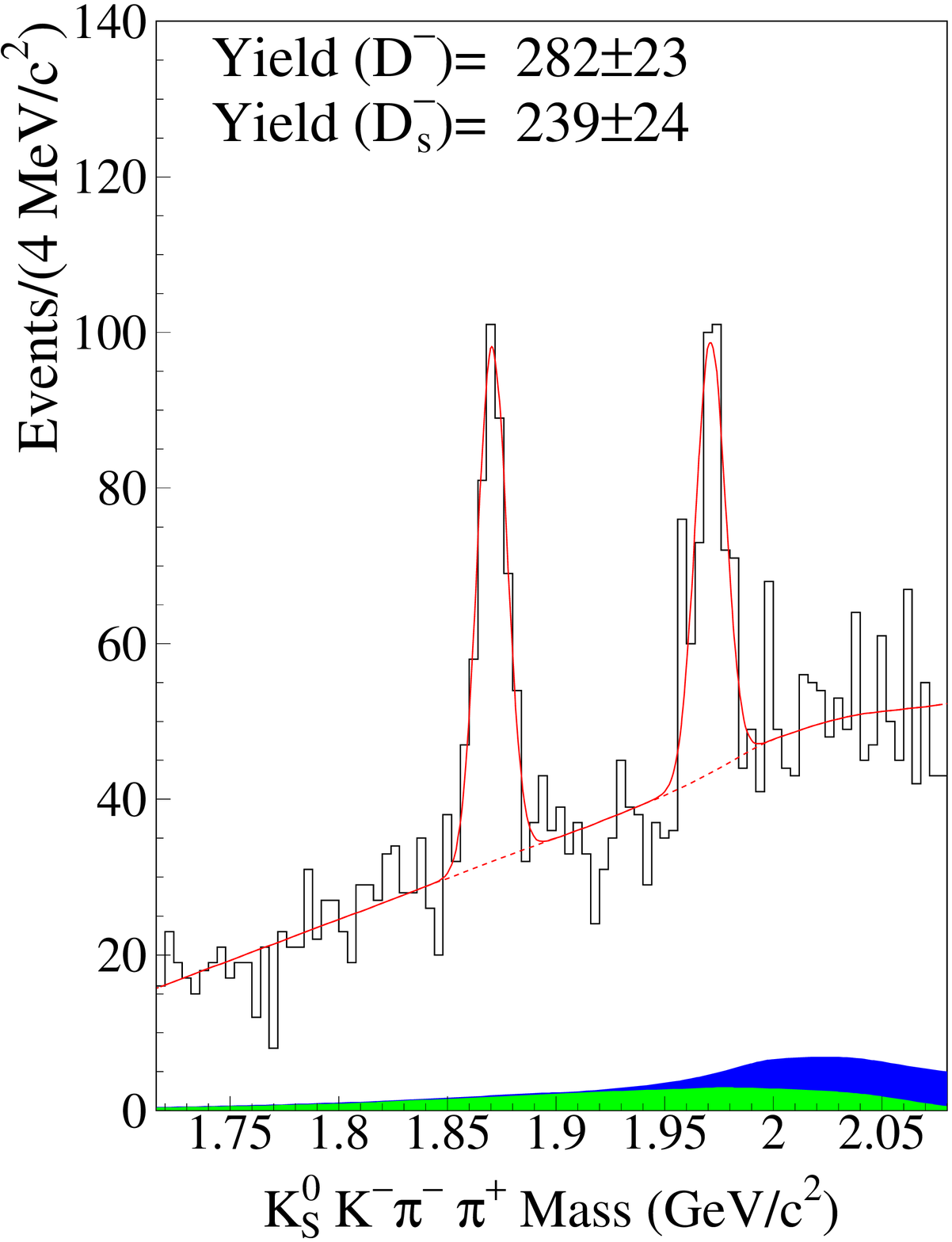}
\end{center}
\caption{Upper plot is an invariant mass plots for the state $K^0_SK^{\pm}\pi^+\pi^-$.
Lower plots are invariant mass plots for $D^+\rightarrow K^0_SK^+\pi^+\pi^-$ (left) and
for  $D^-\rightarrow K^0_SK^-\pi^+\pi^-$ (right).  All plots have an $L/\sigma > 6$ cut.
The broad light shaded region is a reflection from $\Lambda^+_c \to p K^{0}_S\pi^{-}\pi^{+}$, while
the darker shaded region above the $D^+$ mass results from 
$D^+ \to K^0_S\pi^+\pi^{-}\pi^{+}$ where one pion is misidentified as a kaon.}
\label{Fig 4tvio6}
\end{figure}

\begin{figure}[htb!]
\begin{center}
\includegraphics[width=6.5cm]{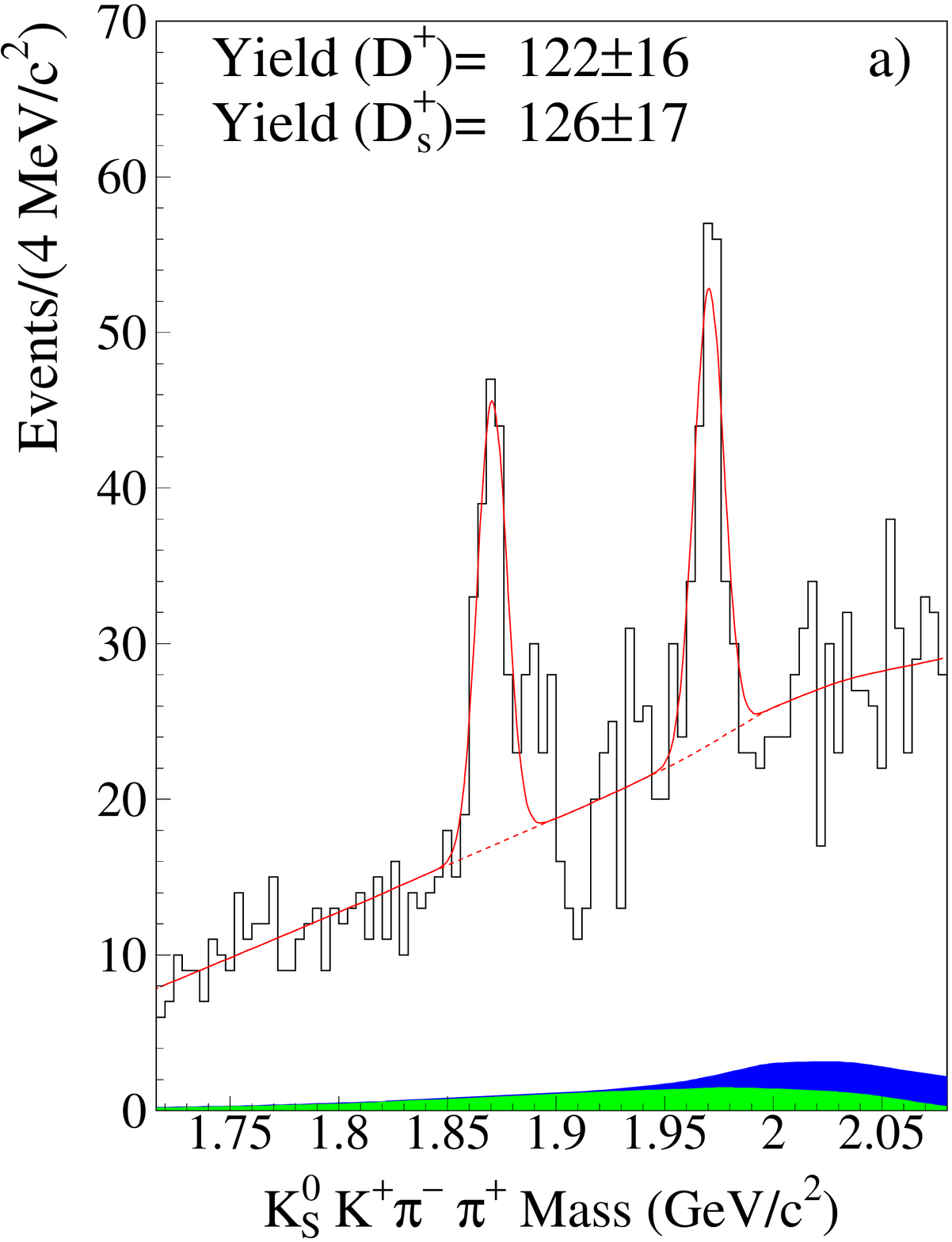}
\includegraphics[width=6.5cm]{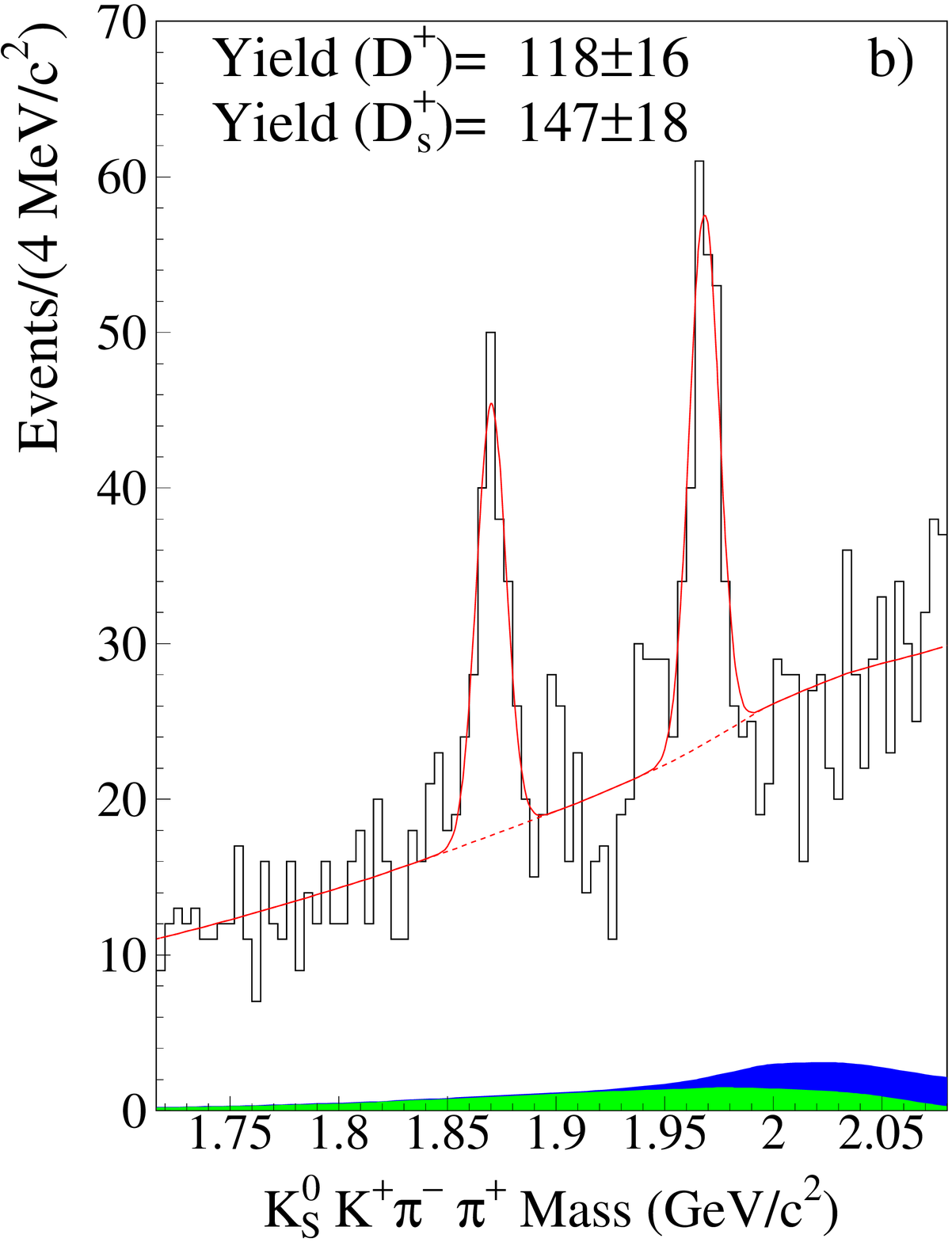}
\includegraphics[width=6.5cm]{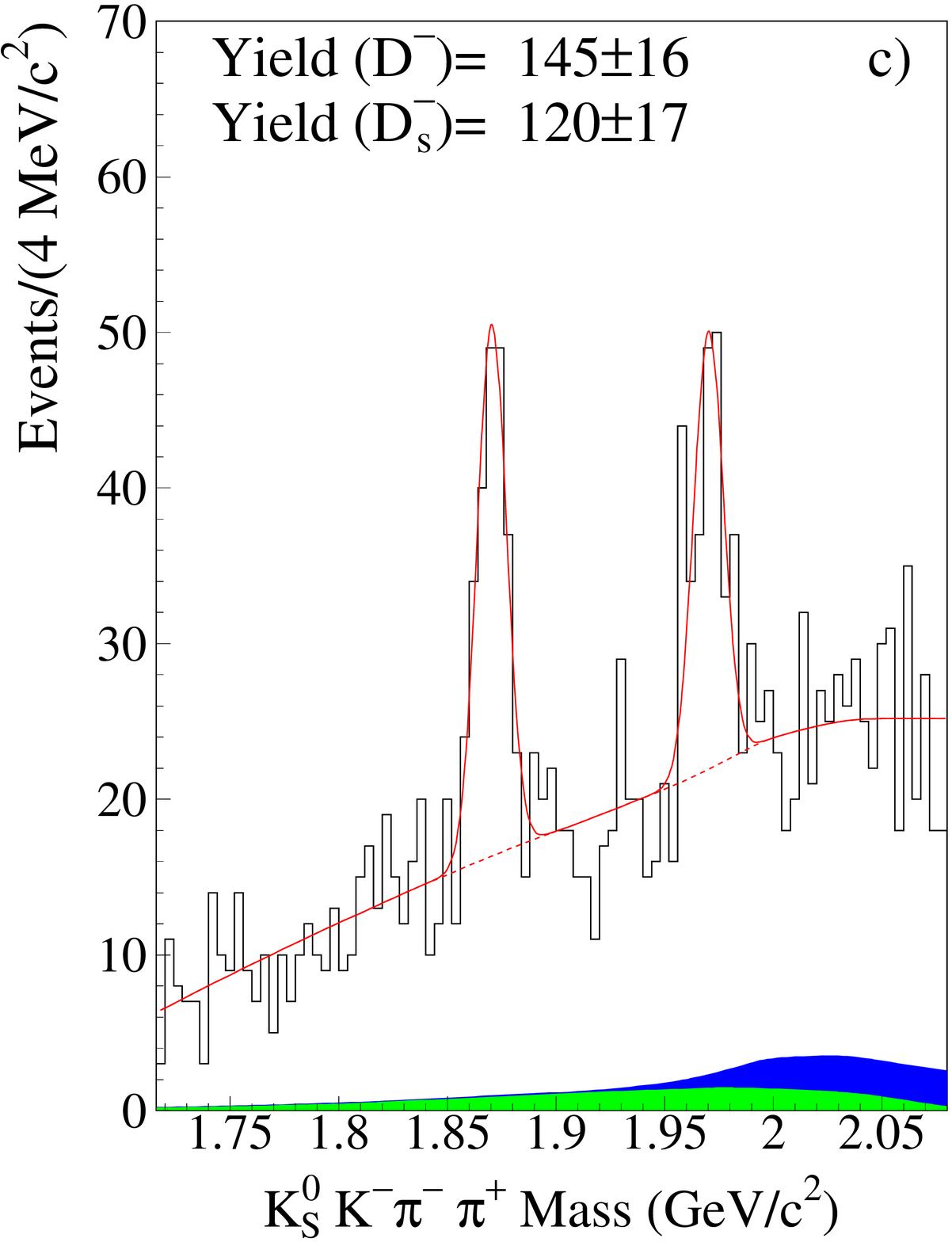}
\includegraphics[width=6.5cm]{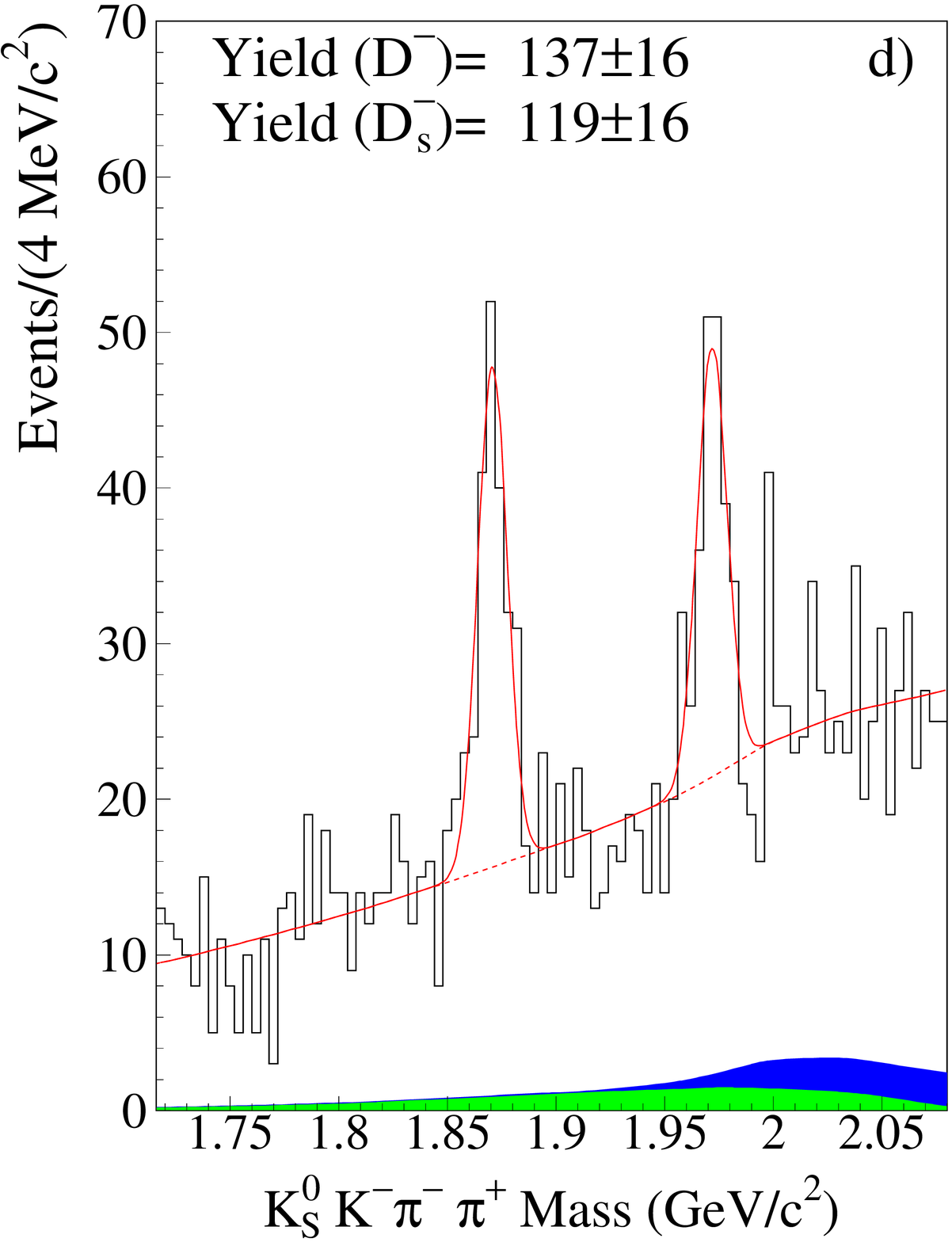}
\end{center}
\caption{Upper plots are invariant mass plots for $D^+\rightarrow K^0_SK^+\pi^+\pi^-$ for $C_T > 0$
in Fig. 5a and for $C_T < 0$ in Fig. 5b. Lower plots are invariant mass plots for 
$D^-\rightarrow K^0_SK^-\pi^+\pi^-$ for $C_T > 0$ in Fig. 5c and for $C_T < 0$ in Fig. 5d. All plots 
have an $L/\sigma > 6$ cut. The shaded regions shown in
each plot are explained in the text and in Fig. 4.}
\end{figure}

\begin{figure}[htb!]
\begin{center}
\includegraphics[width=9.5cm]{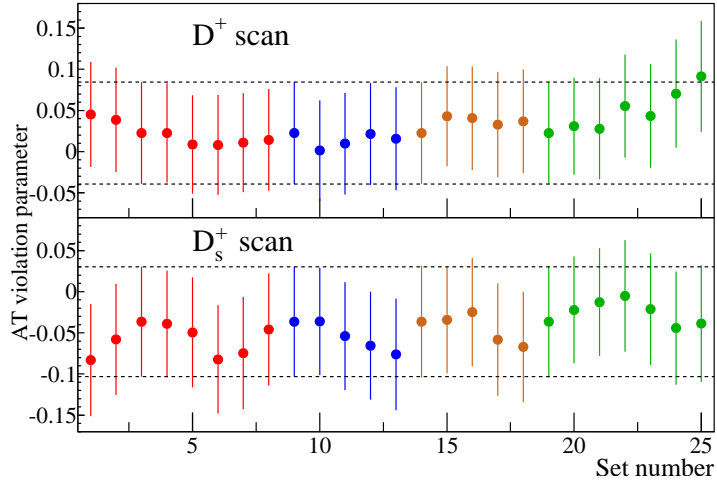}
\end{center}
\caption{\emph{T}-violating asymmetry $A_\mathrm{Tviol}$
versus several sets of cuts. We varied the
$\sigma _{L}$ from $4$ to $12$ (8 points), \emph{Iso2} from $10^{-2}$ to $10^{-6}$ (5 points),
confidence level of the secondary vertex from
$1\%$ to $5\%$ (5 points),
$\Delta _{K}$ from $2$ to $5$ (7 points).
The dashed lines show the quoted $A_\mathrm{Tviol}$ asymmetry $\pm 1 \sigma$.}
\label{Fig 6tvio6}
\end{figure}

\end{document}